\documentclass[12pt]{article}
\usepackage[abs]{overpic}
\usepackage{amssymb, amsmath}
\usepackage{graphicx}
\usepackage{setspace}
\usepackage{cite}
\usepackage{subfig,tikz}

\newcommand{\td}{\text{d}}

\def\be{\begin{equation}}
\def\ee{\end{equation}}
\def\bea{\begin{eqnarray}}
\def\eea{\end{eqnarray}}

\begin{document}

\setcounter{equation}{0}
\numberwithin{equation}{section}

\begin{titlepage}

\vspace*{20mm}
\begin{center}
{\Large \bf Comments on Black Holes in Bubbling Spacetimes}

\vspace*{15mm}
\vspace*{1mm}
 Gary T. Horowitz$^a$,  Hari K. Kunduri$^b$ and James Lucietti$^c$
\vspace*{1cm}
\let\thefootnote\relax\footnote{gary@physics.ucsb.edu, hkkunduri@mun.ca, j.lucietti@ed.ac.uk}

\small{$^a$Department of Physics, University of California\\
Santa Barbara, CA 93106, USA \\
$^b$Department of Mathematics and Statistics, Memorial University of Newfoundland \\ St John's NL A1C 5S7, Canada \\
$^c$School of Mathematics and Maxwell Institute for Mathematical Sciences, \\    University of Edinburgh, King's Buildings, Edinburgh, EH9 3FD, UK
}

\vspace*{1cm}

\end{center}
\begin{abstract}
In five-dimensional minimal supergravity, there are spherical black holes with nontrivial topology outside the horizon which have the same conserved charges at infinity as the BMPV solution. We show that some of these black holes have greater entropy than the BMPV solution. These spacetimes are all asymptotically flat, stationary, and supersymmetric. We also show that there is a limit in which the black hole shrinks to zero size and the solution becomes
a nonsingular ``bubbling" geometry. Thus, these solutions provide explicit analytic examples of placing black holes inside solitons. 
\end{abstract}

\end{titlepage}

%\onehalfspacing
\begin{spacing}{1.2}
\section{Introduction}

Over the past couple of decades, it has become clear that black holes in more than four spacetime dimensions are much less constrained than their four dimensional counterparts. In particular, it is no longer true that stationary black holes are uniquely specified by a few conserved charges at infinity. Some of this nonuniqueness is due to the fact that black hole horizons can have nontrivial topology. The five-dimensional black ring~\cite{Emparan:2001wn} is perhaps the most famous example. A less familiar cause of nonuniqueness is nontrivial topology outside the horizon.

To illustrate this, consider five-dimensional minimal supergravity. This theory admits an asymptotically flat, supersymmetric  black hole with $S^3$ horizon and trivial topology outside ~\cite{Breckenridge:1996is}.  This BMPV black hole is a two-parameter family of solutions characterized by their charge $Q$ and equal angular momenta $J_\psi$ in the two orthogonal planes.\footnote{We will use Euler angles $(\psi, \phi)$ on $S^3$, so for BMPV $J_\phi=0$.} The solution has a regular black hole horizon if
${6 \sqrt{3}\pi} J_\psi^2 < Q^3  $. This theory also has a large class of stationary, asymptotically flat, supersymmetric solutions with no horizons and nontrivial topology (see~\cite{Bena:2007kg} for a review). These ``bubbling" geometries have nontrivial $S^2$'s supported by magnetic flux. Although they are usually studied as candidate nonsingular microstates for a black hole, one can add extremal black holes to these geometries while keeping the solution stationary and supersymmetric. This creates a large class of new spherical black hole solutions.  It was shown in~\cite{Kunduri:2014iga} that the resulting black holes can have the same conserved charges as the  BMPV solution,  providing the first example of continuous non-uniqueness within the class of spherical black holes (supersymmetric or otherwise).

In this note we will show that when $J_\psi$ is close to the BMPV upper bound, the black holes with nontrivial topology outside the horizon can have greater entropy than the BMPV black hole. We will also show that these black holes can exceed the BMPV upper bound on  $J_\psi$. The latter fact is perhaps not surprising since there is structure outside the horizon at larger radius which can carry some of the angular momentum. This is perhaps analogous to the fact that a   black ring  can carry much more angular momentum than a spherical black hole. The first fact, however,  is more surprising. Shortly after the groundbreaking work by Strominger and Vafa~\cite{Strominger:1996sh}, the entropy of the BMPV black holes was reproduced by counting BPS microstates of string theory with the same charges $(Q,J_\psi)$ at weak coupling ~\cite{Breckenridge:1996is}. At the time, the BMPV black hole was the only one known with these charges, so it seemed like a perfect agreement. Now that
we have new solutions with greater entropy, further arguments are needed to understand why the original counting of states reproduces the BMPV entropy. We will discuss this in section 5.

This is not the first time that black hole solutions have been found with the same charges as the BMPV solution and greater entropy. Although a single supersymmetric black ring~\cite{Elvang:2004rt}  cannot have the same charges as BMPV, two concentric supersymmetric black rings can, and sometimes have greater entropy \cite{Gauntlett:2004wh}. (This is a precursor to the four dimensional entropy enigma \cite{Denef:2007vg}.) This phenomenon also occurs for a bound state of two spherical spinning black holes \cite{Crichigno:2016lac}.  If one focusses on the near horizon geometry of BMPV, there are other asymptotically AdS black holes with the same charges and more entropy \cite{Bena:2011zw}.   However, we believe the solutions discussed here are the first examples of asymptotically flat, single horizon black holes with the same charges but greater entropy  than BMPV.\footnote{The recently constructed three-parameter family of black lens solutions~\cite{Kunduri:2014kja} cannot have the same charges as BMPV. This also seems to be the case for the black lens solutions subsequently constructed in~\cite{Tomizawa:2016kjh}.}

We will start with a four parameter family of black holes with a single nontrivial $S^2$ outside the horizon. Setting $J_\phi = 0$ yields a three parameter subset with the same charges as BMPV.  The area of the black hole vanishes along a surface in this parameter space, which marks the  boundary of the physically interesting solutions. Along most of this boundary, the geometry is singular. However there is a set of measure zero where the spacetime is nonsingular and reduces to the original bubbling geometry. 
Near these regular points, one can view the solution as adding a black hole to a soliton.
% a noninteracting mixture of a soliton and a small  extremal black hole. To maximize the entropy, all the angular momentum is carried by the soliton and the black hole is not rotating in this limit. 
There are many previous examples of placing black holes inside solitons~\cite{Volkov:1990,Lee:1991vy,Basu:2010uz,Dias:2011ss} but they are usually only approximate solutions or constructed numerically. Here, as a result of supersymmetry, we have a simple analytic form of the solutions for any size  black hole. (For another recent example, see \cite{Cano:2017qrq}.)

\section{Black hole and bubble spacetime in five dimensions}

Five dimensional minimal supergravity is described by the action 
\be
S = \frac{1}{16\pi G} \int d^5 x \left[\sqrt{-g} (R - F_{mn} F^{mn}) - \frac{2}{3\sqrt 3} \epsilon^{mnpqr}A_m F_{np} F_{qr}\right]
\ee
This theory admits an asymptotically flat, supersymmetric  black hole with $S^3$ horizon and a 2-cycle $C$, or `bubble', outside the horizon~\cite{Kunduri:2014iga}. 
 Here we present the solution in a simpler parameterisation which allows for a more explicit analysis.  Its construction and regularity analysis proceeds in exactly the same fashion as in~\cite{Kunduri:2014iga}, so we will be brief and summarise the main results.

Supersymmetric solutions to minimal supergravity with a Gibbons-Hawking base take the general form
\bea
\td s^2 &=& -f^2 (\td t + \omega_\psi (\td \psi+ \chi) +\hat{\omega})^2 + f^{-1} [ H^{-1} (\td \psi+ \chi)^2 + H (\td r^2+ r^2 \td \Omega_2^2)] \\
A&=& \frac{\sqrt{3}}{2} \left[  f (\td t+ \omega_\psi (\td \psi+ \chi) +\hat{\omega})  - KH^{-1} (\td \psi+ \chi) - \xi \right]
\eea
where $\td \Omega_2^2= \td \theta^2+\sin^2 \theta \td \phi^2$ and  %$(r, \theta, \phi)$ are spherical coordinates on $\mathbb{R}^3$, 
the functions 
\be
f^{-1} = L + H^{-1} K^2, \qquad  \omega_\psi = H^{-2} K^3+ \tfrac{3}{2} H^{-1}KL+M
\ee
are determined by  harmonic functions $H,K,L,M$ on $\mathbb{R}^3$ and the 1-forms  $\chi, \xi, \hat{\omega}$ on $\mathbb{R}^3$ are fixed by these up to quadratures~\cite{Gauntlett:2002nw}.  We take a `3-centred' Gibbons-Hawking base with
\bea
H = \frac{1}{r} - \frac{1}{r_1}+ \frac{1}{r_2}, \qquad \chi =\left(  \cos \theta- \frac{r \cos \theta-a_1}{r_1}+\frac{r \cos \theta-a_2}{r_2} \right) \td \phi
\eea
where $r_1= \sqrt{r^2+ a_1^2 -2a_1 r \cos \theta}$ and $r_2= \sqrt{r^2+ a_2^2 -2a_2 r \cos \theta}$ are the Euclidean distances from the centres and we assume $0<a_1<a_2$. The remaining data are given by
\bea
K &=& \frac{k_1}{r_1}+\frac{k_2}{r_2}, \qquad L = 1+ \frac{\ell_0}{r}+ \frac{k^2_1}{r_1}-\frac{k^2_2}{r_2}, \\ 
M&=& - \frac{3}{2} (k_1+k_2) + \frac{m_0}{r}+ \frac{k^3_1}{2r_1}+\frac{k^3_2}{2r_2} \\
 \xi &=&\left(  - k_1  \frac{r \cos \theta-a_1}{r_1} - k_2 \frac{r \cos \theta-a_2}{r_2} \right)  \td \phi \\
 \hat{\omega} &=& \left( \frac{1}{4 a_1 r} \left[ (k_1^3+2m_0- 3k_1 \ell_0) \left( r_1 + \frac{r^2-a_1^2}{r_1} \right) + 3 r( 2k_1+k_2)  \left( r_1 - \frac{r^2-a_1^2}{r_1} \right) \right] \nonumber \right. \\ &&+ \frac{1}{4 a_2 r} \left[ (k_2^3-2m_0- 3k_2 \ell_0) \left( r_2 + \frac{r^2-a_2^2}{r_2} \right) -3r k_1  \left( r_2 - \frac{r^2-a_2^2}{r_2} \right) \right]  \nonumber \\
&&- \left. \frac{(k_1+k_2)^3(a_1a_2+ r^2- (a_1+a_2)r \cos \theta)}{2(a_2-a_1) r_1 r_2}   + \frac{3}{2} (k_1+k_2) \cos \theta + c \right) \td \phi
\eea
where $(\ell_0, k_1, k_2, m_0, c)$ are constants.

The solution is asymptotically flat $\mathbb{R}^{1,4}$ provided $0<\psi<4\pi$ and $c$ is chosen such that $\hat{\omega} = O(r^{-1})$ as $r\to \infty$. Then, setting $r = \rho^2/4$, as $\rho \to \infty$
\be
\td s^2 \sim - \td t^2+ \td \rho^2 + \tfrac{1}{4}\rho^2 [ (\td \psi + \cos \theta \td \phi)^2 + \td \Omega^2_2]
\ee
with subleading terms $O(\rho^{-2})$.
%\begin{equation}
%c = \frac{3}{2}\left[\frac{\ell_0 k_1}{a_1} + \frac{\ell_0 k_2}{a_2} + \frac{\ell_2 k_1 - \ell_1 k_2}{a_1 - a_2}\right] - \frac{m_0 + m_1}{a_1} + \frac{m_0 - m_2}{a_2} - \frac{m_1 + m_2}{a_1-a_2}
%\end{equation}
The solution is smooth at the `centres' $r_1=0$ and $r_2=0$ if $0<\psi<4\pi$ and the following constraints on the parameters are satisfied
\bea
(\omega_\psi)_{r_1=0}=0 , \qquad (\omega_\psi)_{r_2=0}=0  \label{constraints}
\eea
Lorentzian signature at the centres also requires the inequalities
\bea
f^{-1}_{r_1=0} = \frac{\ell_0- k_1^2+a_1}{a_1} - \frac{(k_1+k_2)^2}{a_2-a_1}<0, \qquad  f^{-1}_{r_2=0} = \frac{\ell_0 - k_2^2 +a_2}{a_2} + \frac{(k_1+k_2)^2}{a_2-a_1}>0 \label{ineq1}
\eea
Near these centres $t=$ constant defines spatial hypersurfaces which approach the origin of $\mathbb{R}^4$. 

The centre $r=0$ is a regular horizon if
\be
\ell_0>0, \qquad \ell_0^3-m_0^2>0   \label{ineqH}
\ee
This can be established by introducing new coordinates $(v, r, \psi', \theta, \phi)$ 
\be
\td t = \td v+ \left( \frac{A_0}{r^2}+ \frac{A_1}{r} \right) \td r, \qquad \td \psi = d\psi'+ \frac{B_0}{r} \td r
\ee
and choosing constants $A_0, A_1, B_0$ such that the metric and its inverse are analytic at $r=0$.\footnote{We find $A_0^2= \ell_0^3-m_0^2$ and $B_0 = A_0m_0/(\ell_0^3-m_0^2)$ with $A_0<0 (>0)$ corresponding to the future (past) horizon.  $A_1$ is a more complicated constant. The Maxwell field is then also analytic at the horizon.} The near-horizon geometry, obtained by the scaling limit $(v,r)\to (v /\epsilon, \epsilon r)$, $\epsilon \to 0$,
 depends only on $(\ell_0, m_0)$ and is given by
\bea\label{eq:nh}
\td s^2_{\text{NH}} &=&- \frac{r^2}{\ell_0^2} \td v^2\pm \frac{2\ell_0}{\sqrt{\ell_0^3- m_0^2}} \td v \td r - \frac{2 m_0 r}{\ell_0^2} \td v (\td \psi' + \cos \theta \td \phi) \nonumber  \\ &+& \left( \ell_0 - \frac{m_0^2}{\ell^2_0} \right) (\td \psi' + \cos \theta \td \phi)^2 + \ell_0 \td \Omega_2^2  \\
F_{\text{NH}} &=& \frac{\sqrt{3}}{2} \td \left[ \frac{r \td v}{\ell_0} +\frac{m_0}{\ell_0} (\td \psi' + \cos \theta \td \phi) \right]
\eea
This is globally isometric to that of the BMPV black hole.
 Spatial cross-sections of the horizon are of $S^3$ topology and have area
 % induced on spatial cross-sections of the horizon is
%\be
%\td s_{H}^2 = \left( \ell_0 - \frac{m_0^2}{\ell^2_0} \right) (\td \psi' + \cos \theta \td \phi)^2 + \ell_0 \td \Omega_2^2
%\ee
%so the area of a cross-section is
\be
A_H= 16 \pi^2 \sqrt{\ell_0^3-m_0^2}
\ee
The above conditions imply  $k_1+k_2 \neq 0  $\footnote{If $k_1+k_2=0$  the constraints (\ref{constraints}) imply $k_1=m_0=0$; in this case (\ref{ineq1}) implies $\ell_0<0$ which is incompatible with a regular horizon. }. This allows us to solve the constraints (\ref{constraints})   uniquely for $(\ell_0, m_0)$. (The solution is not illuminating and will not be given here.)
%\bea
%\ell_0 = -\frac{ a_1 (2k_1+k_2) +k_1 (-a_2+k_2(k_1+k_2))}{k_1+k_2}
%&& m_0= \frac{3 a_1^2 k_2 (2 k_1 + k_2) + a_2 k_1^2 (-3 a_2 + k_1^2 + 4 k_1 k_2 + 3 k_2^2) + 
% a_1 (3 a_2 (k_1^2 - 2 k_1 k_2 - k_2^2) + 
  %  k_2^2 (3 k_1^2 + 4 k_1 k_2 + k_2^2))}{2 (a_1 - a_2) (k_1 + k_2)}
%\eea

The spacetime outside the horizon $r>0$ is smooth if $K^2+HL>0$ and stably causal if $g^{tt}<0$. We have verified numerically that these conditions are satisfied  for a large set of coordinate/parameter values in the special case studied in the next section, provided the above inequalities between the parameters are obeyed. The solution is thus parameterised by the four  constants $(k_1, k_2, a_1, a_2)$ subject to the above inequalities. 

The space outside the horizon has non-trivial topology. Curves between the centres $r_1=0$ and $r_2=0$ correspond to 2-cycles $C$, whereas curves between $r=0$ and $r_1=0$ correspond to non-contractible 2-discs $D$ which end on the horizon. The $z$-axis splits into intervals $I_-=(-\infty, 0),\; I_D=(0, a_1), \;I_C=(a_1, a_2), \; I_+=(a_2, \infty)$, on which different linear combinations of the $U(1)^2$-Killing fields $v_{\pm} = \partial_\phi \pm \partial_\psi$ vanish. The rod diagram \cite{Hollands:2007aj} is given in Figure 1.\footnote{A uniqueness theorem for non-extremal black holes of this type can be found in~\cite{Armas:2014gga}.}
\vskip .5cm
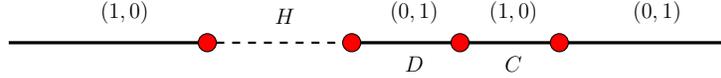
\begin{figure}[h!]
\centering
\subfloat{
\begin{tikzpicture}[scale=1.2, every node/.style={scale=0.7}]
\draw[very thick](-4,0)--(-1.9,0)node[black,left=1.4cm,above=.2cm]{$(1,0)$};
\draw[thick,dashed](-1.7,0)--(-0.3,0)node[black,left=1.1cm,above=.2cm]{$H$};
\draw[very thick](-0.1,0)--(0.9,0)node[black,left=.7cm,above=.2cm]{$(0,1)$} node[black,left=.7cm,below=.1cm]{$D$};
\draw[very thick](1.1,0)--(2.0,0)node[black,left=.7cm,above=.2cm]{$(1,0)$} node[black,left=.7cm, below=.1cm]{$C$};
\draw[very thick](2.2,0)--(4,0)node[black,left=1.4cm,above=.2cm]{$(0,1)$};
\draw[fill=red] (-1.8,0) circle [radius=.1] node[black,font=\large,below=.1cm]{};
\draw[fill=red] (-.2,0) circle [radius=.1] node[black,font=\large,below=.1cm]{};
\draw[fill=red] (1.0,0) circle [radius=.1] node[black,font=\large,below=.1cm]{};
\draw[fill=red] (2.1,0) circle [radius=.1] node[black,font=\large,below=.1cm]{};
\end{tikzpicture}}
\caption{Rod diagram of black hole and bubble spacetime. The pair of integers above each interval specifies the combination of $\partial_\psi, \partial_\phi$ which vanishes there in the basis $(v_+, v_-).$}
\end{figure}

\vskip .5cm

The original BMPV solution is recovered by taking $k_1 = k_2 = 0$, and $a_2 \rightarrow a_1$. This removes the nontrivial topology outside the horizon.

The charge, $Q= \frac{1}{4\pi} \int_{S^3_\infty} \star F $, and angular momenta, $ J_i =  \frac{1}{16\pi} \int_{S^3_\infty} \star \td m_i $, (where $m_i$ are the rotational Killing fields  $ \partial_\psi$ and $ \partial_\phi$)  of the black hole and bubble spacetime are 
\bea
&&Q = 2\sqrt{3}  \pi (\ell_0+  2k_1 (k_1+k_2)) \\
&&J_\psi = \pi( 3(k_1+k_2)(\ell_0 +  k_1(2k_1+k_2))+2m_0)\\
&& J_\phi =3 \pi( a_1 (2k_1+k_2) - a_2 k_1)
\eea 
The mass is determined by the BPS relation $M = \sqrt{3} Q/2$,
and the `dipoles' are\footnote{Unlike the case of a black ring, the dipoles cannot always be expressed as a flux integral. Indeed, $\Pi[D] \equiv \frac{1}{4\pi} \int_D F=q[C]- \frac{m_0}{\ell_0}$ receives a contribution from the horizon. However, $\Pi[C]\equiv \frac{1}{4\pi} \int_C F = q[D]$.}
\bea
q[D] \equiv - \frac{1}{2} v_-^i \Phi_i|_{I_D}=  -\frac{\sqrt{3}}{2}(k_1 + k_2)\;, \qquad q[C] \equiv - \frac{1}{2} v_+^i \Phi_i|_{I_C} = \frac{\sqrt{3}}{2}k_1
\eea
where $\Phi_i$ are magnetic potentials defined by $\nabla_b \Phi_i =   F_{ab} m_i^a$ which vanish at infinity~\cite{Kunduri:2013vka}. 
These five physical quantities are related by the constraint
\begin{equation}
J_\phi = q[D] Q  + \frac{8\pi}{\sqrt{3}} q[D] q[C] \left(q[D] - q[C]\right)
\end{equation}
The area in terms of the physical quantities is
\begin{equation}
A_H = 8 \pi^2\left[\frac{1}{6 \sqrt{3} \pi^3}\left(Q + \frac{16\pi}{\sqrt{3}} q[C] q[D]\right)^3 - \left(\frac{J_\psi + J_\phi}{\pi} + \frac{16}{\sqrt{3}} q[D] q[C]^2\right)^2\right]^{1/2}
\end{equation}
This expression will be the main object of our study below.

By choosing different boundary conditions at $r=0$,  the above family of solutions corresponds to the soliton spacetime with two bubbles found in\cite{Bena:2005va}. This is achieved by imposing that the solution at the centre $r=0$ is smooth, and that near this centre, $t=$ constant defines spatial hypersurfaces which approach the origin of $\mathbb{R}^4$. This requires that $\ell_0=m_0=0$ together with $(\omega_\psi)_{r=0}$=0 and
\bea
%&&(\omega_\psi)_{r=0}= \frac{k_1^3}{2a_1}+ \frac{k_2^3}{2a_2} - \frac{ 3(k_1+k_2)}{2}=0 \\
f^{-1}_{r=0} = 1+ \frac{k_1^2}{a_1}- \frac{k_2^2}{a_2}>0
\eea
In fact, it can be shown that $(\omega_\psi)_{r=0}- (\omega_\psi)_{r_1=0}+ (\omega_\psi)_{r_2=0}=0$ so that $(\omega_\psi)_{r=0}=0$ does not impose any further constraint.  Thus the soliton spacetime is parameterised by $(k_1, k_2, a_1, a_2)$ subject to the constraints (\ref{constraints}) with $\ell_0=m_0=0$, and hence is a 2-parameter family of solutions.

\section{Equal angular momentum phase space}

Interestingly, it was observed in~\cite{Kunduri:2014iga} that there exist black hole and bubble spacetimes with identical global charges to the known BMPV black hole, providing the first example of continuous non-uniqueness within the class of spherical black holes. 
The angular momentum of the BMPV black hole is
\be
J_\phi=0   \label{Jphi0}
\ee
with respect to the Euler angles $(\psi,\phi)$ on the $S^3$ at infinity. It is convenient to express the remaining physical quantities in units of $Q$ (i.e. mass). We define the dimensionless angular momentum and area
\be
\eta \equiv \sqrt{6\pi \sqrt{3}}  \frac{|J_\psi |}{Q^{3/2}} , \qquad a_H \equiv \sqrt{ \frac{3 \sqrt{3} }{32\pi }} \  \frac{A_H}{Q^{3/2}}  \;   \label{etaarea}
 \ee
For the BMPV black hole solution one simply has
\be
0\leq \eta_{\text{BMPV}} <1, \qquad a_{\text{BMPV}} = \sqrt{1- \eta^2}  \; .
\ee
We will now derive the analogous phase space for the black hole and bubble solution discussed above.

For the black hole and bubble solution the angular momentum constraint (\ref{Jphi0}) is
\be
k_2 = - \frac{(2a_1-a_2) k_1}{a_1}   \label{equalJ}
\ee
and hence reduces to a three parameter family $(k_1, a_1, a_2)$. Solving the constraints (\ref{constraints}) we find
\be
\ell_0 =  \frac{(2a_1-a_2) k_1^2}{a_1}, \qquad m_0  = \frac{ a_1^2 k_1 (4k_1^2+3a_2)- a_2 (a_1+a_2) k_1^3}{2 a_1^2}
\ee
and hence
\bea
 f^{-1}_{r_1=0} = \frac{a_1^2 - 2(a_2-a_1) k_1^2}{a_1^2} \; ,\qquad f^{-1}_{r_2=0} =  \frac{a_1 a_2 + 2 (a_2-a_1)k_1^2}{a_1a_2}
\eea
Thus the inequalities (\ref{ineq1}) and (\ref{ineqH}) reduce to
\be\label{ineq2}
2a_1 >a_2, \qquad k_1^2>  \frac{a_1^2}{2(a_2-a_1)}
\ee
together with positivity of the area (this is a more complicated expression).
The electric charge is simply
\be
Q = \frac{2 \sqrt{3} \pi a_2 k_1^2 }{a_1}
\ee
The other physical quantities are most conveniently expressed in terms of (\ref{etaarea}) and a dimensionless dipole
\begin{equation}
\nu \equiv \sqrt{ \frac{\pi q[C]^2}{\sqrt{3}Q}}
\end{equation} 
The area as a function of these quantities is
\be\label{a_H}
a_H= \sqrt{(16\nu^2-1)^3 - \left(\eta + 48\sqrt{2} \nu\left(\nu^2 - \tfrac{1}{8}\right)\right)^2}
\ee
To analyse this formula we need to work out the bounds on $(\eta, \nu)$. 

In terms of the parameters of the solution
\bea
\eta = \frac{3 a_1^2 a_2 +2 k_1^2 a_1 a_2+k_1^2(2a_1-a_2)(a_2-a_1) }{ 2 k_1^2 \sqrt{a_1 a_2^3}}\;, \qquad  \nu = \sqrt{\frac{a_1}{8 a_2}} \;.
\eea 
Observe that (\ref{ineq2}) implies $\eta>0$ so these solutions never possess zero angular momentum.
Using (\ref{ineq2}) we also immediately obtain
\begin{equation}\label{ineqnu}
\frac{1}{4} < \nu < \frac{1}{2 \sqrt{2}} \approx 0.354
\end{equation} To obtain the bounds for the angular momentum $\eta$, it is convenient to introduce an auxiliary dimensionless parameter
\begin{equation}\label{ineqy}
y \equiv \frac{2(a_2 - a_1)k_1^2}{a_1^2} > 1
\end{equation} where the inequality follows from \eqref{ineq2}. A computation shows that one can rewrite this as
\be
y = \frac{6(1- 8\nu^2)}{1 + 4\sqrt{2}\eta \nu -40\nu^2  + 128\nu^4}
\ee Positivity of the denominator and the inequality \eqref{ineqy} impose lower and upper bounds on $\eta$ respectively: 
\begin{equation}
 \frac{-1+40\nu^2 -128 \nu^4}{ 4 \sqrt{2} \nu}<\eta<  \frac{5-8\nu^2 - 128 \nu^4}{ 4 \sqrt{2} \nu}
 \end{equation} However, note that positivity of the area \eqref{a_H} also imposes lower and upper bounds on $\eta$. It may be verified that this provides a more stringent upper bound.  The lower bound is more complicated with the one for the area only providing a more stringent bound below $\nu \approx 0.275$.  Thus we deduce,
\bea
&& \eta_{\text{min}}(\nu) < \eta< \eta_{\text{max}}(\nu)   \; , \label{ineqeta} \\
&&\eta_{\text{min}}  = \text{max} \left( \frac{-1+40\nu^2 -128 \nu^4}{ 4 \sqrt{2} \nu},  -(16 \nu^2-1)^{3/2} + 6 \sqrt{2} \nu (1-8 \nu^2) \right) \nonumber  \\
&&\eta_{\text{max}} = (16 \nu^2-1)^{3/2} + 6 \sqrt{2} \nu (1-8 \nu^2)   \nonumber
\eea
where the max is taken over the range (\ref{ineqnu}). 

In summary, we have shown that the parameter space of the black hole and bubble solution  is given by (\ref{ineqnu}) and (\ref{ineqeta}).
This is plotted in Figure 2. 
\begin{figure}[h]
\centering
\begin{overpic}[height=5cm,width=8cm]{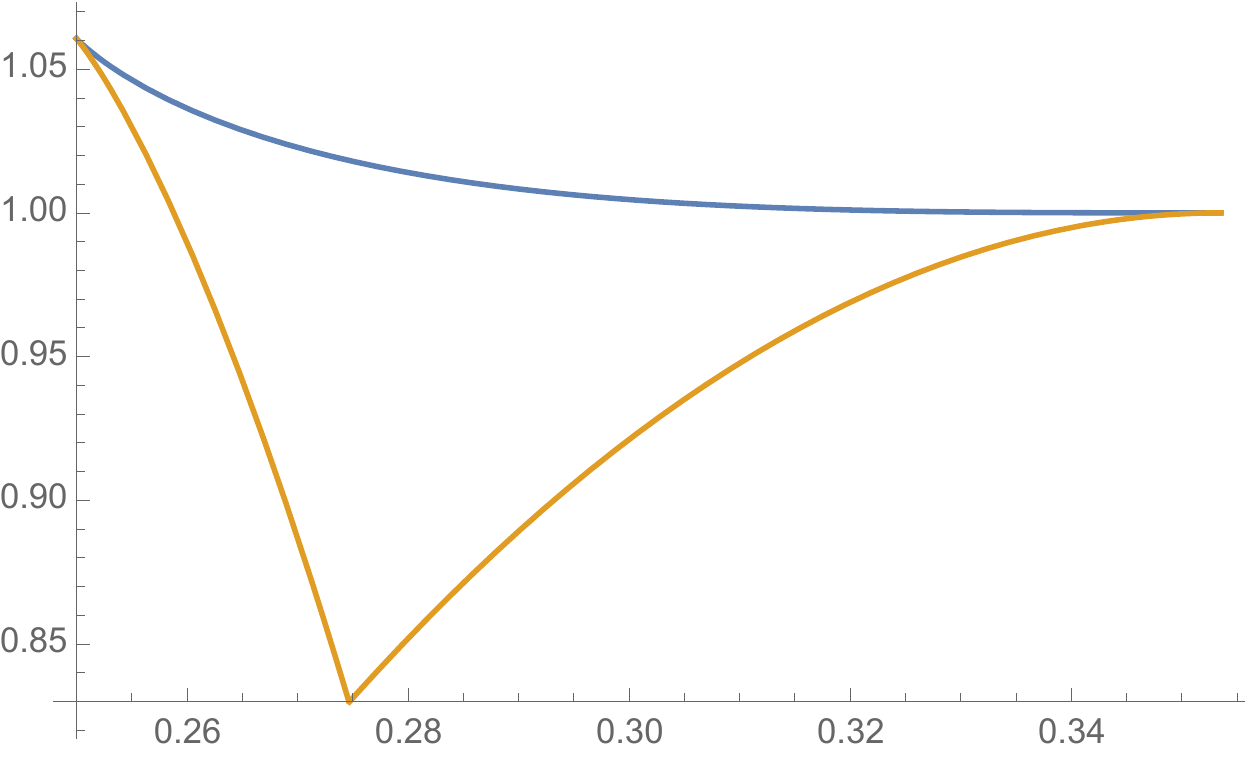}
\put(230,5){$\nu$}
\put(5,140){$\eta$}
\put(100,117){\small{$\underset{\downarrow}{a_H =0}$}}
\put(-2,60){\small{$a_H =0 \rightarrow$}}
\end{overpic}
\caption{$\eta_{\text{max}}$ (blue) and $\eta_{\text{min}}$ (yellow) versus $\nu$ for $1/4 < \nu< 1/(2 \sqrt{2})$. The allowed region is that bounded by the blue and yellow curves. Note that $\eta_{\text{max}}>1$ for all $\nu$. It can been seen that the allowed set of solutions exist on either side of the BMPV bound $\eta=1$.
 }
\end{figure}
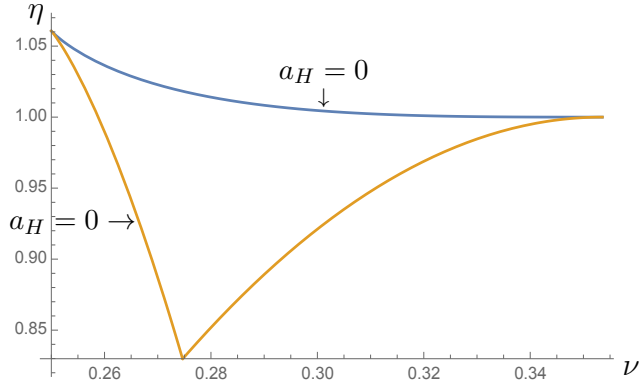 

Let us now consider the smooth soliton solutions discussed at the end of the previous section, with $J_\phi=0$. In terms of the parameters we again must have (\ref{equalJ}). The solution to the constraints (\ref{constraints})  is now\footnote{There is another solution, $k_1=0$; however, this is incompatible with (\ref{ineq1}).}
\be
a_2= 2a_1,  \qquad a_1 = \frac{k_1^2}{3}, \qquad k_2=0 \;.
\ee
Thus the soliton is a 1-parameter family of solutions, parameterised by $k_1 \neq 0$. The physical quantities simplify substantially:
\bea
Q= 4 \sqrt{3} \pi k_1^2, \qquad J_\psi = 6 \pi k_1^3, \qquad q[C]= - q[D] = \frac{\sqrt{3}}{2} k_1
\eea
In terms of the dimensionless quantities:
\be
\eta_s= \frac{3}{2 \sqrt{2}} \approx 1.061, \qquad \nu_s= \frac{1}{4}
\ee
Observe that this corresponds to the point at the lower limit of $\nu$ where $\eta_{\text{max}}= \eta_{\text{min}}$  for the black hole and bubble solution, i.e. the top left hand corner of Figure 2. The rest of the boundary of the allowed black hole region corresponds to naked singularities.

It is interesting to investigate the black hole solution near the soliton point $\eta= \eta_s, \nu = 1/4$.  In fact, the $\eta_{\text{max}}(\nu)$ and $\eta_{\text{min}}(\nu)$ curves are tangent at $\nu = 1/4$. Thus we find that for {\it any} black hole solution in this family, as $\eta \to \eta_s$,
\be
 \nu(\eta) = \frac{1}{4} + \frac{1}{3 \sqrt{2}} ( \eta_s - \eta)+  O(\eta_s - \eta)^2  \label{nearsol}
\ee
and using (\ref{a_H}) this implies
\be
a_H \sim \sqrt{ \frac{128 \sqrt{2}}{ 27} }(\eta_s - \eta)^{3/2} \;.  \label{smallBHarea}
\ee
We will now show one can interpret this as the area of a small nonrotating extremal black hole sitting in the soliton geometry.

The near-horizon geometry is given in (\ref{eq:nh}) and the charge and angular momenta of the corresponding BMPV black hole are\footnote{These can be computed from the near-horizon geometry using appropriate conserved charges defined on the horizon,  see e.g.~\cite{Kunduri:2013ana}.}
\be
\bar{Q} = 2\sqrt{3} \pi \ell_0 , \qquad \bar{J}_\psi= 2\pi m_0
\ee
so the corresponding $\bar{\eta} = m_0/\ell_0^{3/2}$. Then, in terms of the parameters of the full solution
\be
\bar{\eta} = \frac{ \sqrt{2}\eta - 12 \nu( 1-8\nu^2)}{ \sqrt{2} (16 \nu^3-1)^{3/2}}
\ee
Hence, expanding near the soliton point along (\ref{nearsol}) gives 
\be
\bar{\eta}\sim \frac{3 \sqrt{3}}{2^{1/4} 4}  \sqrt{\eta_s - \eta}
\ee
which shows that  the black hole angular momentum vanishes faster than the charge so to leading order, the black hole does not carry angular momentum.  Furthermore, the dimensionless area of the corresponding  extremal black hole (which has $\bar{M} = \sqrt 3 \bar{Q} /2$)  is
\be
\bar{a}_H = \left( \frac{\bar{Q}}{Q} \right)^{3/2} = (16 \nu^2-1)^{3/2} \sim \sqrt{ \frac{128 \sqrt{2}}{ 27} }(\eta_s - \eta)^{3/2} 
\ee
where the last relation is again valid near the soliton point along (\ref{nearsol}).  Thus we find precise agreement with (\ref{smallBHarea}).

This is very similar to previous examples of inserting black holes inside solitons, and agrees with the result of a simple thermodynamic argument. To maximize the entropy,   a noninteracting system of a small black hole and soliton will have all the angular momentum carried by the soliton.  In contrast to most previous examples, however, we now have an explicit analytic form of the solution for any size black hole.  The existence of arbitrarily small black holes implies that the soliton admits static solutions for a charged test particle.  One can check that  a static  test particle with mass $m$ and charge $e$ can indeed be added to the bubbling geometry, but only if $m = \sqrt 3 e /2$. In other words, only if the test particle is also BPS.

\section{Comparison with the BMPV black hole}

Now we will compare the BMPV solution to the black hole and bubble solution in more detail.  In particular, we are interested in when the area of the black hole and bubble solution is greater than (or equal to) the area of the BMPV black hole, so $a_H \geq \sqrt{1-\eta^2}$. Using our explicit formula (\ref{a_H}), this condition is equivalent to
\be
\eta \geq \eta_{\text{crit}} \equiv \frac{1+20 \nu^2 - 32 \nu^4}{6 \sqrt{2} \nu}
\ee
The curve $\eta_{\text{crit}}(\nu)$ is plotted in Figure 3. It can be seen that $\eta_{\text{crit}}$ is very close to the  BMPV bound $\eta=1$ across the whole range of solutions.
\begin{figure}[h]
\centering
\begin{overpic}[height=5cm,width=8cm]{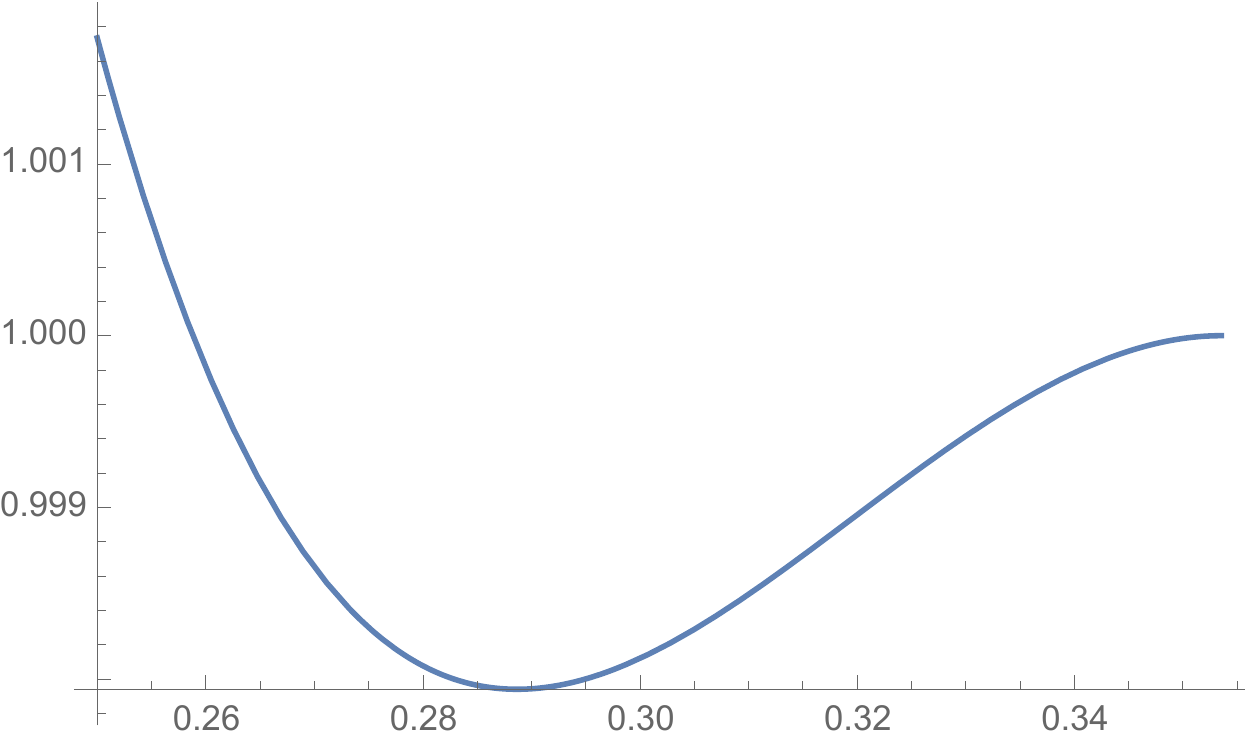}
\put(230,5){$\nu$}
\put(-5,135){$\eta_{\text{crit}}$}
\end{overpic}
\caption{$\eta_{\text{crit}}$ versus $\nu$ for $1/4 < \nu< 1/(2 \sqrt{2})$.}
\end{figure} 

\begin{figure}[h!]
\centering
\begin{overpic}[height=5cm,width=8cm]{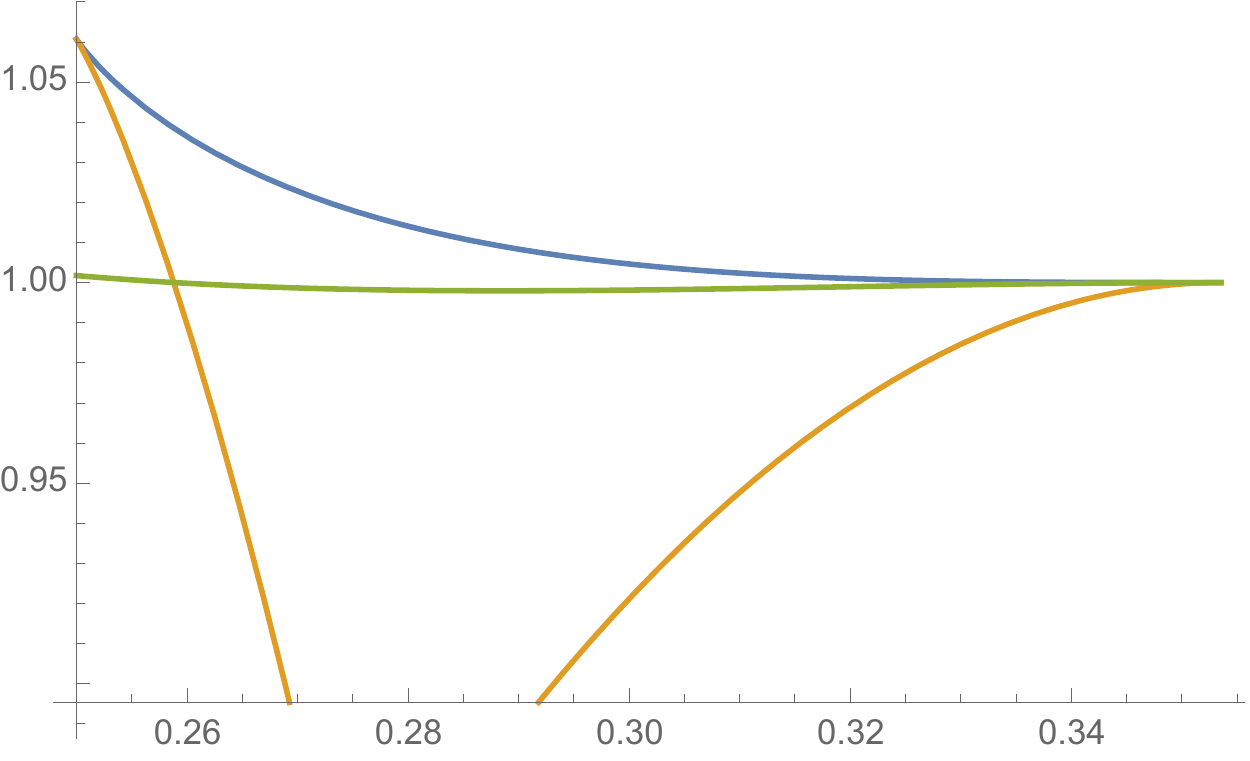}
\put(230,5){$\nu$}
\put(0,140){$\eta$}
\end{overpic}
\caption{$\eta_{\text{max}}$ (blue), $\eta_{\text{min}}$ (yellow) and $\eta_{\text{crit}}$ (green) versus $\nu$ for $1/4 < \nu< 1/(2 \sqrt{2})$.}
\end{figure}

Figure 4 compares $\eta_{\text{crit}}$ with both $\eta_{\text{max}}$ and  $\eta_{\text{min}}$. One can see that $\eta_{\text{crit}} < \eta_{\text{max}}$ for all $\nu$ in the allowed range.  On the other hand, $\eta_{\text{crit}} > \eta_{\text{min}}$  if and only if $\nu >  \tfrac{1}{2} \sqrt{2 - \sqrt{3}} \approx 0.259$. In fact, at the cross-over point $\nu = \tfrac{1}{2} \sqrt{2 - \sqrt{3}} $, it is easily checked that  $\eta_{\text{crit}} = \eta_{\text{min}}=1$. 
 Furthermore, $\eta_{\text{min}}>1$ for all 
 %$1/4< \nu < \tfrac{1}{2} \sqrt{2 - \sqrt{3}} $,
 smaller $\nu$, so there are no corresponding BMPV solutions in this region of phase space. However for 
 %$\tfrac{1}{2} \sqrt{2 - \sqrt{3}} < \nu < 1/(2 \sqrt{2})$, 
 larger $\nu$, we deduce there are two phases of solutions; one with $\eta_{\text{min}} < \eta < \eta_{\text{crit}}<1$ which has lower area than the corresponding BMPV solution, and a second phase with $\eta_{\text{crit}} < \eta< \eta_{\text{max}}$ which for $\eta<1$ coexists with BMPV and has greater area.  
 
Thus we have shown that for 
\be
\frac{1}{2} \sqrt{2 - \sqrt{3}} < \nu < \frac{1}{2 \sqrt{2}}  \label{nurange}
\ee
there is a band of solutions $\eta_{\text{crit}} < \eta< 1$ which have the same conserved charges as the BMPV black hole but {\it greater} entropy.  The area of the two solutions in the region of overlap is plotted in Figure 5.
\begin{figure}[h!]
\centering
\begin{overpic}[height=5cm,width=8cm]{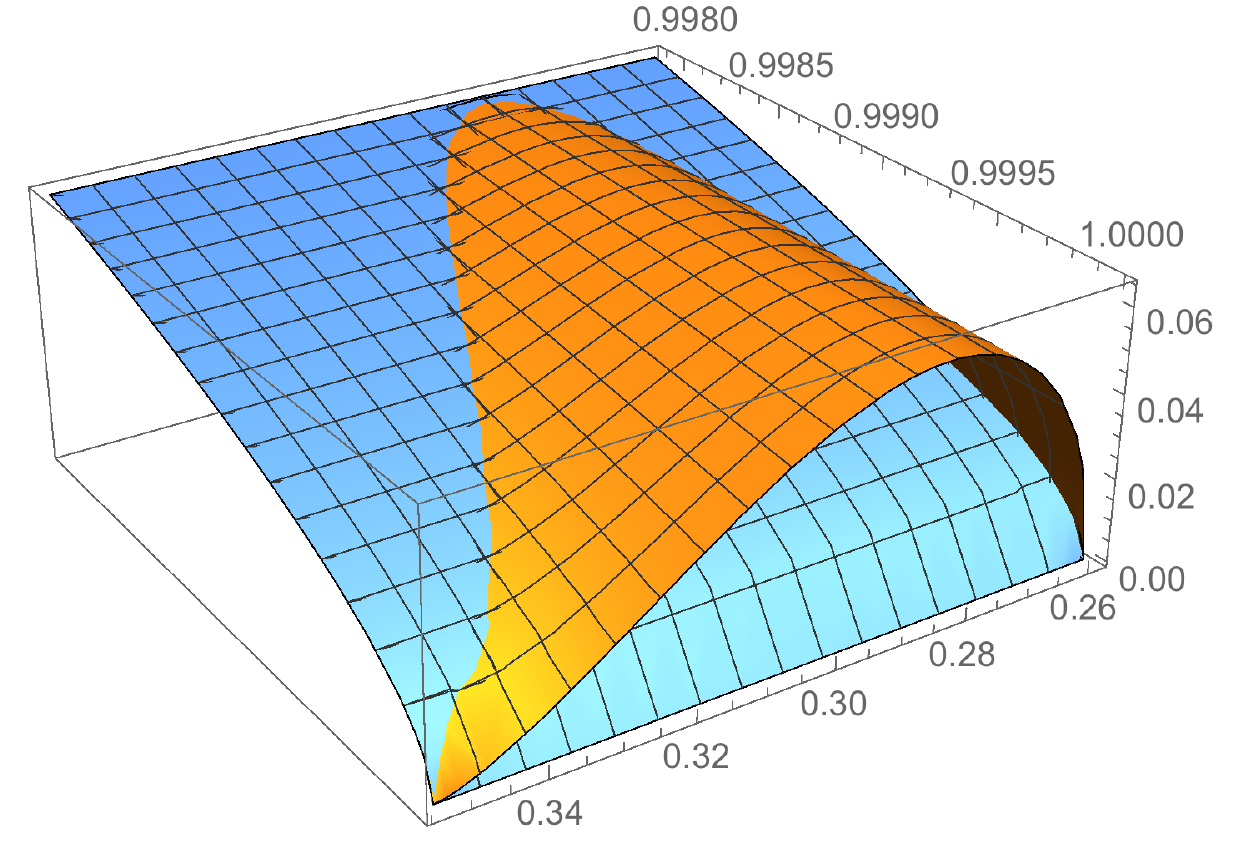}
\put(150,10){$\nu$}
\put(180,125){$\eta$}
\put(225,70){$a_H$}
\end{overpic}
\caption{$a_H$ versus $(\nu, \eta)$ for $\tfrac{1}{2} \sqrt{2 - \sqrt{3}}< \nu< 1/(2 \sqrt{2})$ and $0.998<\eta<1$. The blue surface is the BMPV solution and the orange one is the black hole and bubble solution.}
\end{figure} 

It is interesting to find the maximum entropy state for fixed $\eta$. For $\eta \lesssim 0.998$ the BMPV solution dominates the black hole and bubble solution. However, as can be seen from Figure 5, for $\eta \gtrsim 0.998$ the black hole and bubble solution dominates in the range (\ref{nurange}).  It is clear from the figure that in this region, for a certain value of $\nu=\nu_*(\eta)$, the entropy of the black hole and bubble solution is maximised. Determining $\nu_*(\eta)$ requires finding the appropriate root of $\partial_\nu a_H=0$ (a quintic in $\nu$). Fortunately, since the region of interest $0.998 \lesssim \eta <1$ is very close to one, we may determine $\nu_*(\eta)$ to good accuracy by expanding in $(1-\eta)$.  Indeed, we find
\bea
\nu_*(\eta) \approx 0.284+ 2.025 (1-\eta)  \\
a_{\text{max}}(\eta) \approx 0.059+ 2.404 (1-\eta)
\eea
In contrast, for BMPV near $\eta=1$ we have $a_{\text{BMPV}} \approx \sqrt{2(1-\eta)}$.

\section{Discussion}

We have studied a four-parameter family of black hole solutions with a topologically  nontrivial $S^2$-cycle outside the horizon.  We found that there is a three-dimensional subset with the same charges as the BMPV black hole, some of which contain greater entropy. This might be viewed as a ``single black hole entropy enigma".

From the gravitational standpoint, there is a natural explanation for this phenomenon. The new black holes only have greater entropy  when the angular momentum $J_\psi$ is close to the BMPV upper bound $J(Q) \equiv (Q^3/6\sqrt 3 \pi)^{1/2}$. However the entropy of a BMPV black hole vanishes as $J_\psi$ approaches $J(Q)$.  The new solutions have structure outside the horizon which can carry angular momentum. So when the total angular momentum approaches $J(Q)$, the remaining angular momentum carried by the black hole is less than this and hence the entropy remains nonzero.

 One might object that the configuration outside the black hole carries charge as well as angular momentum, so it is a priori possible that the charge of the black hole would also be reduced leaving the BMPV bound unaffected. However we have seen that the new family of black hole solutions can have $J_\psi > J(Q)$, and have a near horizon geometry that is the same as BMPV, with shifted parameters. This shows that the nontrivial topology outside the horizon carries relatively more angular momentum than charge so if the total quantities satisfy $J_\psi = J(Q)$, the black hole itself carries  $J_\psi < J(Q)$ and has a nonzero entropy. 

The question remains why the original counting of microstates  \cite{Breckenridge:1996is}  gave the correct entropy for the BMPV black hole and not  one of these new solutions. Even though $J_\psi$ and $Q$ are quantized in string theory, in the limit of large $Q$  many discrete values would lie inside the region where the new black holes have greater entropy.  The original counting involved computing a certain index (the elliptic genus)  in weakly coupled string theory and extrapolating to strong coupling. It was always possible that this index undercounted the number of BPS states. A recent construction \cite{Haghighat:2015ega} has indeed  found weakly coupled BPS states with $J_\psi > J(Q)$, but although the number of such states is exponentially large, it cannot explain the entropy of the macroscopic black holes discussed here. 

One might think that a possible explanation for the original agreement is that since the bubbling geometries have nontrivial topology, they are nonperturbative solutions that cannot be seen in string perturbation theory. So the original counting of microstates in Minkowski spacetime could not include  black holes sitting in these spacetimes. However, it has been argued that as one decreases the string coupling, the bubbles reduce to wrapped branes which {\it can} be seen at weak coupling \cite{Balasubramanian:2006gi}.  So  either the  index calculation undercounts the number of BPS states, or there are more complicated BPS bound states of branes and strings that are not included in the D-brane system that was originally studied.

We can try to get a deeper understanding using holography. The new black hole solutions can be lifted  to six dimensions and  the asymptotically flat region removed by taking a decoupling limit. The resulting spacetime is asymptotically $AdS_3 \times S^3$, and one can ask what are the dual CFT states that  they correspond to. Unfortunately, even without adding black holes, the CFT dual of the bubbling geometries are not yet known  \cite{Skenderis:2008qn,Giusto:2015dfa}.

We should note that it has recently been argued that the bubbling geometries are all nonlinearly unstable \cite{Eperon:2016cdd}, in the sense that adding a small finite amount of energy will change the solution by a large amount. The likely endpoint of this instability is a state with string scale curvature and large stringy corrections to supergravity \cite{Marolf:2016nwu}. The same instability probably applies to the black hole solutions discussed here.

Finally, we have examined just the simplest example of a black hole with nontrivial topology outside the horizon. Many more examples could be constructed and explored. For example, a supersymmetric black hole with $n$ nontrivial 2-cycles outside the horizon could be constructed in the Gibbons-Hawking class by taking harmonic functions with $n+2$ centres.  As argued above, structure outside the horizon can carry angular momentum and hence extra 2-cycles could decrease the proportion of angular momentum carried by the black hole. 
This suggests that, as the total angular momentum approaches $J(Q)$, adding 2-cycles outside the black hole could further increase the entropy.  By continuity, this argument also suggests that the region of phase space where the entropy dominates over BMPV would increase (i.e. the lower bound on $\eta$ would decrease). It would be interesting to investigate this in more detail.  

%\vskip 1cm
%\centerline{\bf Acknowledgements} 
%\vskip .5cm
\section*{Acknowledgements}

It is a pleasure to thank D. Marolf  and A. Strominger for discussions. GTH is supported in part by NSF Grant PHY-1504541. HKK is supported by NSERC Discovery Grant 418537-2012. JL is supported in part by the Science and Technology Facilities Council (STFC) [ST/L000458/1].

\end{spacing}

\end{document}